\documentstyle[12pt,epsf,rotate]{article}

\newcommand{\AmS}{{\protect\the\textfont2
  A\kern-.1667em\lower.5ex\hbox{M}\kern-.125emS}}
\newcommand{\be}{\begin{equation}}
\newcommand{\ee}{\end{equation}}
\newcommand{\bea}{\begin{eqnarray}}
\newcommand{\eea}{\end{eqnarray}}
\newcommand{\bi}{\bibitem}
\newcommand{\nn}{\nonumber}

\newcommand{\complex}{{{\rm I} \kern -.59em {\rm C}}}

\catcode`\@=11
\def\lsim{\mathrel{\mathpalette\@versim<}}
\def\gsim{\mathrel{\mathpalette\@versim>}}
\def\@versim#1#2{\vcenter{\offinterlineskip
\ialign{$\m@th#1\hfil##\hfil$\crcr#2\crcr\sim\crcr } }}
\catcode`\@=12

\begin{document}

\begin{titlepage}
  \renewcommand{\thefootnote}{\fnsymbol{footnote}}
  \vspace*{-7\baselineskip}
  \begin{flushright}
    \begin{tabular}{l@{}}
      NTUA-59/96\\
      IFUNAM-FT97-4\\  
       hep-ph/9704218
    \end{tabular}
  \end{flushright}

  \vskip 0pt plus 0.4fill

  \begin{center}
    \textbf{\LARGE Reduction of Couplings and Finiteness in Realistic
    Supersymmetric GUTs \footnote{Presented by G. Zoupanos at the
    Buckow Symposium 1996} }

  \end{center}

  \vskip 0pt plus 0.2fill

  \begin{center}
    {\large
    J. Kubo%
    \\}
    \textit{Physics Dept.\\
      Faculty of Natural Sciences\\
      Kanazawa University\\
      920-11 Kanazawa, Japan}\\
    \vspace{1ex}
    {\large
    M. Mondrag\'on%
    \footnote{Partially supported by the PAPIIT project IN110296}
    \\}
    \textit{
      Instituto de F\1sica, UNAM\\
      Apdo. Postal 20-364\\
      M\'exico 1000 D.F., M\'exico}\\
      and\\
    {\large
    G. Zoupanos%
    \footnote{Partially supported by the E.C.
    projects CHRX-CT93-0319, ERBFMRXCT960090, and the Greek project
    PENED/1170}
    \\} 
    \textit{ 
      Physics Deptartment\\
      Nat. Technical University\\
      157 80 Zografou, Athens, Greece}\\
    
    \vskip 1ex plus 0.3fill


    \vskip 1ex plus 0.7fill

    \textbf{Abstract}
  \end{center}

\noindent
Reduction of couplings in supersymmetric GUTs is achieved by searching
for renormalization  group invariant (RGI) relations among couplings
which hold beyond the unification scale.  Finiteness is due to the
fact that there exist RGI relations among couplings that guarantee the
vanishing of the $\beta$-functions of a $N=1$ supersymmetric GUT even
to all orders in perturbation theory.  Of particular interest are the
relations among gauge and Yukawa couplings which lead to very
interesting predictions of the top quark mass.    

  \vskip 0pt plus 2fill

  \setcounter{footnote}{0}

\end{titlepage}
        

\section{Introduction}

The essence of all theoretical efforts in Elementary Particle Physics
(EPP) is to understand the present day free parameters of the Standard
Model (SM) in terms of a few fundamental ones, i.e. to achieve {\em
  reduction of couplings}.  The pathology of the plethora of free
parameters is well known.  It is deeply connected to the presence of
{\em infinities} at the quantum level.  The renormalization program
can remove the infinities by introducing counterterms, but only at
the cost of leaving the corresponding terms as free parameters.   The
traditional way of reducing the number of parameters is to assume 
that the world is more {\em symmetric} at higher scales.  A celebrated
example along this line is the minimal $SU(5)$, whose gauge sector
provided us with a testable prediction for one of the low energy gauge
couplings.  In fact nowadays LEP data suggest that the minimal
$SU(5)$ GUT should be even more symmetric, specifically $N=1$
supersymmetric.  However, requiring more symmetry, e.g. constructing
GUTs based on groups of higher rank, such as $SO(10),~E_6,~E_7,~E_8$,
is well known that does not increase the predictive power of these
models.  An extreme case from this point of view are superstrings or
superbranes,
which have huge symmetries but no predictions for the SM parameters.

A natural extension of the GUT idea is to find a way to relate the
gauge and Yukawa sectors of a theory, that is to achieve Gauge-Yukawa
Unification (GYU).  A symmetry which naturally relates the two sectors
is supersymmetry, in particular $N=2$ supersymmetry.  It turns out,
however, that $N=2$ supersymmetric theories have serious
phenomenological problems due to light mirror fermions.  Also, in
superstring theories and in composite models there exist relations
among the gauge and Yukawa couplings, but both kind of theories have
phenomenological problems.

In a series of papers \cite{kmz,mondragon2,kmtz,kmtz2,kmz-pert} we
have proposed 
another way to relate the gauge and Yukawa sectors of a theory. It is
based on the fact that 
within the framework of a renormalizable field theory, one can find
renormalization group invariant (RGI) relations among parameters
that can improve the calculability and the predictive power of a
theory.  We have considered models in which the GYU is achieved 
using the principles of reduction of couplings
\cite{cheng1,zim1,kubo1} and
finiteness
\cite{kmz},\cite{PW}-\cite{model},\cite{LPS}-\cite{piguet},\cite{al}.
These principles, 
which are formulated in 
perturbation theory, are not explicit symmetry principles, although
they might imply symmetries \cite{sib-app19b}.  The {\em reduction of
  couplings} is based
on the existence of RGI relations among couplings, which preserve 
perturbative renormalizability.  

The principle of {\em finiteness} maybe today requires some
    more motivation in order to be generally accepted.  It is however 
  interesting  to note that in the old days the general feeling was
    quite different.  Probably the well known Dirac's phrase that
    ``...divergences are hidden under the carpet'' is representative
    of the views of that time.

In recent years we have a more relaxed attitude towards divergencies.
Most theorists believe that the divergencies are signals of the
existence of a higher scale, where new degrees of freedom are
excited.  Parenthetically one should note that even accepting this
philosophy everybody is bothered by the presence of quadratic
divergencies in a theory, since their presence implies that physics at
one scale is very sensitive to unknown physics at higher scales.  This
was and still remains the main reason for introducing supersymmetric
theories in EPP, since they provide us with the only known perturbative
examples of theories free of quadratic divergences.

Returning to the argument that divergences reflect the existence of
new physics at higher scales, we are naturally led to the conclusion
that beyond unification scale, i.e. when all interactions have been
taken into account in a unified scheme, the theory should be
completely finite.  In fact this is one of the main motivations and 
aims of string, non-commutative geometry and quantum group theories,
which include also gravity in the unification of the interactions.

In our attempts here we are restricted in unifying only the known
gauge interactions, and it is interesting to point out that {\em finiteness
  does not require gravity}.  Finiteness is based on the fact that it
is possible to find RGI relations among couplings that keep finiteness
in perturbation theory, even to all orders.

Applying the principles of reduction of couplings and finiteness one
can relate the gauge and Yukawa couplings and therefore improve the
predictive power of a model.

In the following we present the principles of reduction of couplings
and finiteness as well as their application in realistic
supersymmetric GUTs.  In addition we present some hints on the deeper
connections among the reduction of couplings, supersymmetry and
finiteness which certainly requires further exploration.

\section{Reduction of Couplings by the RGI Method}

Let us next briefly outline the idea of reduction of couplings.  
Any RGI relation among couplings 
(which does not depend on the renormalization
scale $\mu$ explicitly) can be expressed,
in the implicit form $\Phi (g_1,\cdots,g_A) ~=~\mbox{const.}$,
which
has to satisfy the partial differential equation (PDE)
\bea
\mu\,\frac{d \Phi}{d \mu} &=& {\vec \nabla}\cdot {\vec \beta} ~=~ 
\sum_{a=1}^{A} 
\,\beta_{a}\,\frac{\partial \Phi}{\partial g_{a}}~=~0~,
\eea
where $\beta_a$ is the $\beta$-function of $g_a$.
This PDE is equivalent
to a set of ordinary differential equations, 
the so-called reduction equations (REs) \cite{zim1},
\bea
\beta_{g} \,\frac{d g_{a}}{d g} &=&\beta_{a}~,~a=1,\cdots,A~,
\label{redeq}
\eea
where $g$ and $\beta_{g}$ are the primary 
coupling and its $\beta$-function,
and the counting on $a$ does not include $g$.
Since maximally ($A-1$) independent 
RGI ``constraints'' 
in the $A$-dimensional space of couplings
can be imposed by the $\Phi_a$'s, one could in principle
express all the couplings in terms of 
a single coupling $g$.
 The strongest requirement is to demand
 power series solutions to the REs,
\bea
g_{a} &=& \sum_{n=0}\rho_{a}^{(n)}\,g^{2n+1}~,
\label{powerser}
\eea
which formally preserve perturbative renormalizability.
Remarkably, the 
uniqueness of such power series solutions
can be decided already at the one-loop level \cite{zim1}.
To illustrate this, let us assume that the $\beta$-functions
have the form
\bea
\beta_{a} &=&\frac{1}{16 \pi^2}\{
\sum_{b,c,d\neq g}\beta^{(1)\,bcd}_{a}g_b g_c g_d\nn\\
&+&\sum_{b\neq g}\beta^{(1)\,b}_{a}g_b g^2\}+\cdots~,\nn\\
\beta_{g} &=&\frac{1}{16 \pi^2}\beta^{(1)}_{g}g^3+
\cdots~,
\eea
where $\cdots$ stands for higher order terms, and 
$ \beta^{(1)\,bcd}_{a}$'s are symmetric in $b,c,d$.
 We then assume that
the $\rho_{a}^{(n)}$'s with $n\leq r$
have been uniquely determined. To obtain $\rho_{a}^{(r+1)}$'s,
we insert the power series (\ref{powerser}) into the REs (\ref{redeq})
and collect terms of  
$O(g^{2r+3})$ to find
\bea
\sum_{d\neq g}M(r)_{a}^{d}\,\rho_{d}^{(r+1)} &=&
\mbox{lower order quantities}~,\nn
\eea
where the r.h.s. is known by assumption, and
\bea
M(r)_{a}^{d} &=&3\sum_{b,c\neq g}\,\beta^{(1)\,bcd}_{a}\,\rho_{b}^{(1)}\,
\rho_{c}^{(1)}+\beta^{(1)\,d}_{a}\nn\\
&-&(2r+1)\,\beta^{(1)}_{g}\,\delta_{a}^{d}~,\label{M}\\
0 &=&\sum_{b,c,d\neq g}\,\beta^{(1)\,bcd}_{a}\,
\rho_{b}^{(1)}\,\rho_{c}^{(1)}\,\rho_{d}^{(1)}\nn\\
&+&\sum_{d\neq g}\beta^{(1)\,d}_{a}\,\rho_{d}^{(1)}
-\beta^{(1)}_{g}\,\rho_{a}^{(1)}~.
\eea
 Therefore,
the $\rho_{a}^{(n)}$'s for all $n > 1$
for a given set of $\rho_{a}^{(1)}$'s can be uniquely determined if
$\det M(n)_{a}^{d} \neq 0$  for all $n \geq 0$.

It is clear by examining specific examples,
that the various 
couplings in supersymmetric theories have easily the same asymptotic
behaviour.  Therefore searching for a power series solution of the
form (\ref{powerser})
to the REs (\ref{redeq}) is justified. This is not the case in
non-supersymmetric theories.

The above facts lead us to suspect that there is and intimate
connection among the requirement of reduction of couplings and
supersymmetry which still waits to be uncovered.  The connection
becomes more clear by examining the following example.  

Consider an $SU(N)$ gauge theory with the following matter content:
$\phi^{i}({\bf N})$ and $\hat{\phi}_{i}(\overline{\bf N})$
are complex scalars, 
$\psi^{i}({\bf N})$ and $\hat{\psi}_{i}(\overline{\bf N})$
are left-handed Weyl spinor,
and $\lambda^a (a=1,\dots,N^2-1)$ is
right-handed Weyl spinor in the adjoint representation of $SU(N)$.

The Lagrangian is:
\bea
&&{\cal L} = -\frac{1}{4}F_{\mu\nu}^{a}F^{a \,\mu\nu}+
i \sqrt{2} \{~g_Y\overline{\psi}\lambda^a T^a \phi\nn\\
&&-\hat{g}_Y\overline{\hat{\psi}}\lambda^a T^a \hat{\phi}
+\mbox{h.c.}~\}-
V(\phi,\overline{\phi}),\\
&&V(\phi,\overline{\phi}) =
\frac{1}{4}\lambda_1(\phi^i \phi^{*}_{i})^2+
\frac{1}{4}\lambda_2(\hat{\phi}_i \hat{\phi}^{*~i})^2\\
&&+\lambda_3(\phi^i \phi^{*}_{i})(\hat{\phi}_j \hat{\phi}^{*~j})+
\lambda_4(\phi^i \phi^{*}_{j})
(\hat{\phi}_i \hat{\phi}^{*~j}),\nn
\eea
which is the most general renormalizable form
of dimension four, consistent with the $SU(N)\times SU(N)$ global
symmetry.

Searching for solution (\ref{powerser}) of the REs (\ref{redeq}) we
find in lowest order the following one:
\bea
g_{Y}&=&\hat{g}_{Y}=g~,~\nn\\
\lambda_{1}&=&\lambda_{2}=\frac{N-1}{N}g^2~,~\\
\lambda_{3}&=&\frac{1}{2N}g^2~,~
\lambda_{4}=-\frac{1}{2}g^2~.\nn
\eea
which corresponds to an $N=1$ supersymmetric gauge theory
\footnote{Further details will be given in a future publication
  \cite{future}.}. Clearly the above remarks do not answer the
question of the relation among reduction of couplings and
supersymmetry but rather try to trigger the interest for further
investigation. 

\section{Reduction of Couplings and N=1 Supersymmetric Gauge Theories}

Let us consider a chiral, anomaly free,
$N=1$ globally supersymmetric
gauge theory based on a group G with gauge coupling
constant $g$. The
superpotential of the theory is given by
\bea
W&=& \frac{1}{2}\,m_{ij} \,\phi_{i}\,\phi_{j}+
\frac{1}{6}\,C_{ijk} \,\phi_{i}\,\phi_{j}\,\phi_{k}~,
\label{supot}
\eea
where $m_{ij}$ and $C_{ijk}$ are gauge invariant tensors and
the matter field $\phi_{i}$ transforms
according to the irreducible representation  $R_{i}$
of the gauge group $G$. 

The $N=1$ non-renormalization theorem \cite{nonre} ensures that
there are no mass
and cubic-interaction-term infinities.
As a result the only surviving possible infinities are
the wave-function renormalization constants
$Z^{j}_{i}$, i.e.,  one infinity
for each field. The one -loop $\beta$-function of the gauge
coupling $g$ is given by \cite{PW}
\bea
\beta^{(1)}_{g}=\frac{d g}{d t} =
\frac{g^3}{16\pi^2}\,\{\,\sum_{i}\,l(R_{i})-3\,C_{2}(G)\,\}~,
\label{betag}
\eea
where $l(R_{i})$ is the Dynkin index of $R_{i}$ and $C_{2}(G)$
 is the
quadratic Casimir of the adjoint representation of the
gauge group $G$. The $\beta$-functions of
$C_{ijk}$,
by virtue of the non-renormalization theorem, are related to the
anomalous dimension matrix $\gamma_{ij}$ of the matter fields
$\phi_{i}$ as:
\be
\beta_{ijk} =
 \frac{d C_{ijk}}{d t}~=~C_{ijl}\,\gamma^{l}_{k}+
 C_{ikl}\,\gamma^{l}_{j}+
 C_{jkl}\,\gamma^{l}_{i}~.
\label{betay}
\ee
At one-loop level $\gamma_{ij}$ is \cite{PW}
\be
\gamma_{ij}^{(1)}=\frac{1}{32\pi^2}\,\{\,
C^{ikl}\,C_{jkl}-2\,g^2\,C_{2}(R_{i})\delta_{ij}\,\},
\label{gamay}
\ee
where $C_{2}(R_{i})$ is the quadratic Casimir of the representation
$R_{i}$, and $C^{ijk}=C_{ijk}^{*}$.
Since
dimensional coupling parameters such as masses  and couplings of cubic
scalar field terms do not influence the asymptotic properties 
 of a theory on which we are interested here, it is
sufficient to take into account only the dimensionless supersymmetric
couplings such as $g$ and $C_{ijk}$.
So we neglect the existence of dimensional parameters, and
assume furthermore that
$C_{ijk}$ are real so that $C_{ijk}^2$ always are positive numbers.
For our purposes, it is
convenient to work with the square of the couplings and to
arrange $C_{ijk}$ in such
a way that they are covered by a single index $i~(i=1,\cdots,n)$:
\bea
\alpha &=& \frac{|g|^2}{4\pi}~,~
\alpha_{i} ~=~ \frac{|g_i|^2}{4\pi}~.
\label{alfas}
\eea
We define 
\bea
\tilde{\alpha}_{i} &\equiv& \frac{\alpha_{i}}{\alpha}~,~i=1,\cdots,n~,
\label{alfat}
\eea
and derive a form for the evolution equations in terms of $\alpha$
and $\tilde \alpha$:
\bea
\alpha \frac{d \tilde{\alpha}_{i}}{d\alpha} =
-\tilde{\alpha}_{i}&+&\frac{\beta_{i}}{\beta}~=~
(\,-1+\frac{\beta^{(1)}_{i}}{\beta^{(1)}}\,)\, \tilde{\alpha}_{i}\nn\\
-\sum_{j,k}\,\frac{\beta^{(1)}_{i,jk}}{\beta^{(1)}}
\,\tilde{\alpha}_{j}\, \tilde{\alpha}_{k}&+&\sum_{r=2}\,
(\frac{\alpha}{\pi})^{r-1}\,\tilde{\beta}^{(r)}_{i}(\tilde{\alpha})~,
\label{RE}
\eea
where $\tilde{\beta}^{(r)}_{i}(\tilde{\alpha})~(r=2,\cdots)$
are power series of $\tilde{\alpha}$'s and can be computed
from the $r$-th loop $\beta$-functions.
Next we search for fixed points $\rho_{i}$ of Eq. (\ref{alfat}) at $ \alpha
= 0$. To this end, we have to solve
\bea
(\,-1+\frac{\beta ^{(1)}_{i}}{\beta ^{(1)}}\,)\, \rho_{i}
-\sum_{j,k}\frac{\beta ^{(1)}_{i,jk}}{\beta ^{(1)}}
\,\rho_{j}\, \rho_{k}&=&0~,
\label{fixpt}
\eea
and assume that the fixed points have the form
\bea
\rho_{i}&=&0~\mbox{for}~ i=1,\cdots,n'~;~\\
&&\rho_{i} ~>0 ~\mbox{for}~i=n'+1,\cdots,n~.\nn
\eea
We then regard $\tilde{\alpha}_{i}$ with $i \leq n'$
 as small
perturbations  to the
undisturbed system which is defined by setting
$\tilde{\alpha}_{i}$  with $i \leq n'$ equal to zero.
As we have seen,
it is possible to verify at the one-loop level \cite{zim1} the
existence of the unique power series solution
\bea
\tilde{\alpha}_{i}&=&\rho_{i}+\sum_{r=2}\rho^{(r)}_{i}\,\alpha^{r-1}~,~\\
&&i=n'+1,\cdots,n~\nn
\label{usol}
\eea
of the reduction equations (\ref{RE}) to all orders in the undisturbed
system. 
These are RGI relations among couplings and keep formally
perturbative renormalizability of the undisturbed system.
So in the undisturbed system there is only {\em one independent}
coupling, the primary coupling $\alpha$.

\subsection{The Minimal Supersymmetric $SU(5)$ Model}

The minimal N=1 supersymmetric $SU(5)$ model is
particularly interesting, 
being the the simplest GUT supported by the LEP data.  Here we will
consider it as an attractive example of a partially reduced model.  Its
particle content is well defined and has the following transformation
properties under SU(5):
three $({\bf \overline{5}}+{\bf 10})$-
supermultiplets which accommodate three fermion families,
one $({\bf 5}+{\bf \overline{5}})$ to describe the two Higgs
supermultiplets appropriate for electroweak symmetry breaking
and a ${\bf 24}$-supermultiplet required to provide the
spontaneous 
symmetry breaking of $SU(5)$ down to
$SU(3)\times SU(2) \times U(1)$.

\begin{table*}
\begin{center}\caption{The predictions of the minimal SUSY
    $SU(5)$}\label{table-afut} 
\vspace{0.4 cm}
\begin{tabular}{|c|c|c|c|c|c|c|c|}
\hline
$m_{\rm SUSY}$ [GeV]
& $g_{t}^{2}/g^2$ & $g_{b}^{2}/g^2$&
$\alpha_{3}(M_Z)$ &
$\tan \beta$  &  $M_{\rm GUT}$ [GeV]
 & $m_{b} $ [GeV]& $m_{t}$ [GeV]
\\ \hline
$300$ & $0.97$& $0.57$ & $0.120$  & $47.7$ & $1.8\times10^{16}$
  & $5.4$  & $179.7 $  \\ \hline
$500$ & $0.97$& $0.57$ & $0.118$  & $47.7$ & $1.39\times10^{16}$
  & $5.3$  & $178.9$  \\ \hline
\end{tabular}
\end{center}
\end{table*}

We require the GYU to occur among the Yukawa couplings of the third
generation and the gauge coupling.  In addition we require the theory
to be completely asimptotically free.

Neglecting the dimensional parameters
and the Yukawa couplings of the first two generations,
the superpotential of the model is exactly given by
\bea
W &=& \frac{1}{2}\,\,g_{t} {\bf 10}_{3}\,
{\bf 10}_{3}\,H+ \sqrt{2}
g_{b}\, \overline{{\bf 5}}_{3}\,{\bf 10}_{3}\, \overline{H}\\
&&+{1\over 3}g_{\lambda}\,({\bf 24})^3+
g_{f}\,\overline{H}\,{\bf 24}\, H~,\nn
\eea
where $H, \overline{H}$ are the ${\bf 5},\overline{{\bf 5}}$-
Higgs supermultiplets and we have suppressed the $SU(5)$
indices.

In the one-loop approximation, the GYU yields
$$g_{t,b}^{2} ~=~\sum_{m,n=1}^{\infty}
\kappa^{(m,n)}_{t,b}~[\frac{g_{\lambda}}{g}]^{2m}
\,[\frac{g_{f}}{g}]^{2n}~g^2$$
($h$ and $f$ are related
to the Higgs couplings). Where 
$[g_{\lambda}/g]^2$ is allowed to vary from
$0$ to $15/7$, while 
$[g_{f}/g]^2$ may vary from $0$ to a maximum which depends
on $[g_{\lambda}/g]^2$ and vanishes at $[g_{\lambda}/g]^2=15/7$.
As a result, we obtain \cite{kmoz2}
\bea
0.97\,g^2 \le &g_{t}^{2}& \le 1.37\,g^2~,\nonumber\\
{}~0.57\,g^2 \le g_{b}^{2}&=&g_{\tau}^{2}  \le 0.97\,g^2~.
\eea

We found \cite{kmoz2} that consistency with proton decay requires
$g_t^2,~g_b^2$ to be very close to the left hand side values in the
inequalities.

Just below the unification scale we would like to obtain the MSSM
$SU(3)\times SU(2)\times U(1)$ and one pair of Higgs doublets, and
assume that all the superpartners 
are degenerate at the supersymmetry breaking scale, where the MSSM
will be broken to the normal SM. 
Then the standard model should  be spontaneously broken down to
$SU(3)\times U(1)_{\rm em}$ due to the VEVs of the two Higgs
$SU(2)$-doublets contained in the ${\bf 5},\overline{{\bf
5}}$-super-multiplets.

One way to obtain the correct low energy theory is to add to
the Lagrangian soft supersymmetry breaking
terms and to arrange
the mass parameters in the superpotential along with
the soft breaking terms so that
the desired symmetry breaking pattern of the original $SU(5)$
is really the preferred one, all the superpartners are
unobservable at present energies,
there is no contradiction with proton decay,
and so forth.
Then we study 
the evolution of the couplings at two loops
respecting all the boundary conditions at $M_{GUT}$.

 In Table 1 we give the predictions for representative
values of $m_{SUSY}$, where we have suppressed the threshold effects
of the superheavy as well as of the MSSM particles.

\section{Finiteness and Reduction of Couplings in Extended
  Supersymmetric Gauge Theories}

In the introduction we have presented arguments suggesting that a truly
unified theory should be finite.  The two main new points that we
would like to advertise in the present and the following two chapters
are a) that there exist finite gauge theories, i.e. unified theories
which are finite without the inclusion of gravity, and b) that there
exists a finite realistic GUT with testable low energy predictions.
Furthermore we would like to present suggestive hints on the deeper
relation among reduction of couplings and finiteness which reopens the
old relevant discussion.

Concerning finiteness in gauge theories it is well known that there
exists a non-renormalization theorem \cite{sw-plb100}, which
guarantees that gauge 
theories based on any gauge group with extended $N=4$ supersymmetry is
finite to all orders.  On the other hand it is also well known that
nobody even dares to attempt the construction of realistic models
based on $N=4$.

The next best candidates gauge theories to become finite are those
having $N=2$ supersymmetry.  The reason is the existence of another
non-renormalization theorem \cite{gs-npb201} stating that their $\beta$-
functions obtain only one-loop contributions.  Therefore for a given
gauge group one could choose, in principle, appropriate matter fields so
as to obtain a vanishing one-loop $\beta$-function and then the
non-renormalization theorem guarantees the finiteness to all orders.
To be more specific, the $\beta$-function of a $N=2$ gauge theory is
\cite{fz-npb79}: 
\be
\beta(g)={2g^3\over 4 \pi^2} (\sum_i T(R_i)-C_2(G)).
\ee
Therefore a $N=2$ gauge theory based on the group $SU(N)$ could be
made finite by introducing $2N$ multiplets transforming according to
the fundamental representation.  In particular searching for finite
$N=2$ GUTs one finds \cite{pw-plb127} that the introduction
of the matter fields in the 
following representations and corresponding multiplicities is
required:
\be 
{\bf SU(5)}:~~~p({\bf 5}+{\bf\bar5});~q({\bf 10}+{\bf\bar{10}});~r({\bf
  15}+{\bf\bar{15}})
\ee
with the multiplicities $p,~q,~,r$ satisfying the constraint
\be
p + 3q + 7r = 10.
\ee
\be
{\bf SO(10)}:~~~p({\bf 10}+{\bf\bar{10}});~~~q({\bf 16}+{\bf\bar{16}})
\ee
with
\be
p+2q = 8.
\ee
\be
{\bf E_(6)}:~~~4({\bf 27}+{\bf\bar{27}})
\ee
Obviously the above GUTs, in order to become finite, require the
presence of multiplets corresponding to mirror particles.

Therefore, the construction of a realistic finite GUT necessitates the
invention of a mechanism which keeps ordinary fermions light and makes
the mirror fermions unobservably heavy.  Due to the lack of such a
mechanism at present, the question of constructing realistic $N=2$
GUTs is postponed but certainly not excluded.

Let us turn next to a discussion of the relation among extended
supersymmetries and the reduction of couplings by examining specific
gauge theories with the particle content of $N=2,4$ pure super
Yang-Mills (SYM) theories (i.e.~without matter fields) but without the
corresponding couplings \cite{sib-app19b,osz-plb147}.

\subsection{Frame theory for $N=2$ SYM}

Consider the following $R$-invariant Lagrangian
\bea
 L &=& -{1\over 4 g^2}F^{\mu\nu}F_{\mu\nu} + {1\over 2}
(D_{\mu}\phi^a)^2 + {1\over 2}(D_{\mu}\pi^a)^2 \nn\\ 
&-&i\bar{\psi}^a\gamma D \psi ^a - i \sqrt{g_1}\epsilon^{abc}
\bar{\psi}(\phi^a+\gamma_5 \pi^b)\psi^c \nn\\
&-& {1\over
  4}g_2(\phi^2+\pi^2)^2 
+{1\over 4}g_3(\phi^a\phi^b+\pi^a\pi^b)^2,
\eea
where we have a supermultiplet $\phi^a$ of real scalar, $\pi^a$ of
pseudoscalar, $A^a_{\mu}$ of  vector field, 
and $\psi^a$ of Dirac spinor, all transforming according to the
adjoint representation of
$SU(2)$.  The Lagrangian has the field content of 
pure $N=2$ SYM-theory in components and allows for a
non-supersymmetric embedding of it.  Searching for power series
solutions (\ref{powerser}) of the REs (\ref{redeq}) one finds two
solutions of the RE with positive classical potential, and uniquely 
determined at all orders.  One which has 
the $N=2$ symmetry at tree level
\be
\rho_1^{(0)}=\rho_2^{(0)}=\rho_3^{(0)}=1,
\ee
and one which does not have the symmetry
\be
\rho_1^{(0)}=1,~~\rho_2^{(0)}={9\over\sqrt{105}},~~\rho_3^{(0)}={7\over\sqrt{10
    5}}.
\ee

\subsection{Frame Theory for $N=4$ SYM}

Consider now the following $N=4$ SYM Lagrangian, which is required to
be $SU(2)\times SU(2)$ rigid invariant for simplicity
\bea
L&=& -{1\over 4 g^2}F^{\mu\nu}F_{\mu\nu} + {1\over 2}
(D_{\mu}\phi^a_i)^2 \nn\\
&+& {1\over 2}(D_{\mu}\pi^a_l)^2  
-{i\over 2}\bar{\psi}^a_K \not \!\!D \psi ^a_K\\ 
&-& {1\over 2} \sqrt{g_1} \epsilon
^{abc}\bar{\psi}^a_K(\alpha^i_{KL}\phi^b_i+\gamma_5\beta^i_{KL}\pi^b_i)
\psi^c_L\nn\\
&-& {1\over 4} g_2(\phi^a_i\phi^a_i+\pi^a_l\pi^a_l)^2 +{1\over 4}
g_2(\phi^a_i\phi^b_i+\pi^a_l\pi^b_l)^2, \nn
\eea
where we have 3 scalar $\phi_i$, 3 pseudoscalars $\pi_i$, 4 Majorana
spinors $\psi _K$, and one vector $A_{\mu}$, all transforming
according to the adjoint representation of $SU(2)$.

Searching again for solutions of the RE (\ref{redeq}) we get two
allowed power series solutions (\ref{powerser}). The first one has
$N=4$ symmetry 
\be
\rho_1^{(0)}=\rho_2^{(0)}=\rho_3^{(0)}=1.
\ee
The second one does not show any symmetry at the tree approximation
\be
\rho_1^{(0)}=1,~~\rho_2^{(0)}=0.7579\dots,~~\rho_3^{(0)}=0.2523\dots
\ee
and it is worth noticing that its $\beta$-functions vanish in the
one-loop order also. 
 
Thus we see that the requirement of reduction of couplings admits a
symmetric solution corresponding to the extended supersymmetric
theories but it is more general, since it has also other solutions.

\section{Finiteness and Reduction of Couplings in $N=1$ SUSY Gauge Theories}

As one can see from Eqs.~(\ref{betag}) and (\ref{gamay}) in Chapter 3,
 all the one-loop $\beta$-functions of the theory vanish if
 $\beta_g^{(1)}$ and $\gamma _{ij}^{(1)}$ vanish, i.e.
\begin{equation}
\sum _i \ell (R_i) = 3 C_2(G) \,,
\label{1st}
\end{equation}

\begin{equation}
C^{ikl} C_{jkl} = 2\delta ^i_j g^2  C_2(R_i)\,,
\label{2nd}
\end{equation}

A very interesting result is that the conditions (\ref{1st},\ref{2nd}) are
necessary and sufficient for finiteness at
the two-loop level \cite{PW}.

In case supersymmetry is broken by soft terms, one-loop finiteness of
the soft sector imposes further constraints on it \cite{soft}.  In
addition, the same set of conditions that are sufficient for one-loop
finiteness of the soft breaking terms render the soft sector of the
theory two-loop finite \cite{jj}.

The one- and two-loop finiteness conditions (\ref{1st},\ref{2nd}) restrict
considerably the possible choices of the irreps. $R_i$ for a given
group $G$ as well as the Yukawa couplings in the superpotential
(\ref{supot}).  Note in particular that the finiteness conditions cannot be
applied to the supersymmetric standard model (SSM), since the presence
of a $U(1)$ gauge group is incompatible with the condition
(\ref{1st}), due to $C_2[U(1)]=0$.  This naturally leads to the
expectation that finiteness should be attained at the grand unified
level only, the SSM being just the corresponding, low-energy,
effective theory.

Another important consequence of one- and two-loop finiteness is that
supersymmetry (most probably) can only be broken by soft breaking
terms.  Indeed, due to the unacceptability of gauge singlets, F-type
spontaneous symmetry breaking \cite{raifer} terms are incompatible
with finiteness, as well as D-type \cite{fayet} spontaneous breaking
which requires the existence of a $U(1)$ gauge group.

A natural question to ask is what happens at higher loop orders.  The
answer is contained in a theorem \cite{LPS} which states the necessary
and sufficient conditions to achieve finiteness at all orders.  Before
we discuss the theorem let us make some introductory remarks.  The
finiteness conditions impose relations between gauge and Yukawa
couplings.  To require such relations which render the couplings
mutually dependent at a given renormalization point is trivial.  What
is not trivial is to guarantee that relations leading to a reduction
of the couplings hold at any renormalization point.  As we have seen,
the necessary, but also sufficient, condition for this to happen is to
require that such relations are solutions to the REs
\be
\beta_g {d C_{ijk}\over dg} = \beta_{ijk}
\label{redeq2}
\ee
and hold at all orders.  As we have seen, remarkably the existence of
all-order power series solutions to (\ref{redeq2}) can be decided at
the one-loop level.

Let us now turn to the all-order finiteness theorem \cite{LPS}, which
states when a $N=1$ supersymmetric gauge theory can become finite to
all orders in the sense of vanishing $\beta$-functions, that is of
physical scale invariance.  It
is based on (a) the structure of the supercurrent in $N=1$ SYM
\cite{fz-npb87,pisi-npb196}, and on 
(b) the non-renormalization properties of $N=1$ chiral anomalies
\cite{LPS,pisi}. 
Details on the proof can be found in refs. \cite{LPS} and further
discussion in refs.~\cite{pisi,LZ,piguet}.  Here, following mostly
ref.~\cite{piguet} we present a comprehensible sketch of the proof.
 
Consider a $N=1$ supersymmetric gauge theory, with simple Lie group
$G$.  The content of this theory is given at the classical level by
the matter supermultiplets $S_i$, which contain a scalar field
$\phi_i$ and a Weyl spinor $\psi_{ia}$, and the vector supermultiplet
$V_a$, which contains a gauge vector field $A_{\mu}^a$ and a gaugino
Weyl spinor $\lambda^a_{\alpha}$. 

Let us first recall certain facts about the theory:

\noindent (1)  A massless $N=1$ supersymmetric theory is invariant 
under a $U(1)$ chiral transformation $R$ under which the various fields 
transform as follows
\bea
A'_{\mu}&=&A_{\mu},~~\lambda '_{\alpha}=\exp({-i\theta})\lambda_{\alpha}\nn\\ 
\phi '&=& \exp({-i{2\over
    3}\theta})\phi,~~\psi_{\alpha}'= \exp({-i{1\over
    3}\theta})\psi_{\alpha},~\cdots
\eea
The corresponding axial Noether current $J^{\mu}_R(x)$ is
\be
J^{\mu}_R(x)=\bar{\lambda}\gamma^{\mu}\gamma^5\lambda + \cdots
\label{noethcurr}
\ee
is conserved classically, while in the quantum case is violated by the
axial anomaly
\be
\partial_{\mu} J^{\mu}_R =
r(\epsilon^{\mu\nu\sigma\rho}F_{\mu\nu}F_{\sigma\rho}+\cdots).
\label{anomaly}
\ee

{}From its known topological origin in ordinary gauge theories
\cite{ag-npb243}, one would expect that the axial vector current
$J^{\mu}_R$ to satisfy the Adler-Bardeen theorem  and
receive corrections only at the one-loop level.  Indeed it has been
shown that the same non-renormalization theorem holds also in
supersymmetric theories \cite{pisi}.  Therefore
\be
r=\hbar \beta_g^{(1)}.
\label{r}
\ee

\noindent (2)  The massless theory we consider is scale invariant at
the classical level and, in general, there is a scale anomaly due to
radiative corrections.  The scale anomaly appears in the trace of the
energy momentum tensor $T_{\mu\nu}$, which is traceless classically.
It has the form
\bea
T^{\mu}_{\mu} &~=~& \beta_g F^{\mu\nu}F_{\mu\nu} +\cdots
\label{Tmm}
\eea

\noindent (3)  Massless, $N=1$ supersymmetric gauge theories are
classically invariant under the supersymmetric extension of the
conformal group -- the superconformal group.  Examining the
superconformal algebra, it can be seen that the subset of
superconformal transformations consisting of translations,
supersymmetry transformations, and axial $R$ transformations is closed
under supersymmetry, i.e. these transformations form a representation
of supersymmetry.  It follows that the conserved currents
corresponding to these transformations make up a supermultiplet
represented by an axial vector superfield called supercurrent
$J$,
\be
J \equiv \{ J'^{\mu}_R, ~Q^{\mu}_{\alpha}, ~T^{\mu}_{\nu} , ... \},
\label{J}
\ee
where $J'^{\mu}_R$ is the current associated to R invariance,
$Q^{\mu}_{\alpha}$ is the one associated to supersymmetry invariance,
and $T^{\mu}_{\nu}$ the one associated to translational invariance
(energy-momentum tensor). 

The anomalies of the R current $J'^{\mu}_R$, the trace
anomalies of the 
supersymmetry current, and the energy-momentum tensor, form also
a second supermultiplet, called the supertrace anomaly
\bea
S &=& \{ Re~ S, ~Im~ S,~S_{\alpha}\} =\nn\\
&& \{T^{\mu}_{\mu},~\partial _{\mu} J'^{\mu}_R,~\sigma^{\mu}_{\alpha
  \dot{\beta}} \bar{Q}^{\dot\beta}_{\mu}~+~\cdots \}
\eea
where $T^{\mu}_{\mu}$ in Eq.(\ref{Tmm}) and
\bea
\partial _{\mu} J'^{\mu}_R &~=~&\beta_g\epsilon^{\mu\nu\sigma\rho}
F_{\mu\nu}F_{\sigma\rho}+\cdots\\ 
\sigma^{\mu}_{\alpha \dot{\beta}} \bar{Q}^{\dot\beta}_{\mu}&~=~&\beta_g
\lambda^{\beta}\sigma^{\mu\nu}_{\alpha\beta}F_{\mu\nu}+\cdots 
\eea

\noindent (4) It is very important to note that 
the Noether current defined in (\ref{noethcurr}) is not the same as the
current associated to R invariance that appears in the
supercurrent 
$J$ in (\ref{J}), but they coincide in the tree approximation. 
So starting from a unique classical Noether current
$J^{\mu}_{R(class)}$,  the Noether
current $J^{\mu}_R$ is defined as the quantum extension of
$J^{\mu}_{R(class)}$ which allows for the
validity of the non-renormalization theorem.  On the other hand
$J'^{\mu}_R$, is defined to belong to the supercurrent $J$,
together with the energy-momentum tensor.  The two requirements
cannot be fulfilled by a single current operator at the same time.

Although the Noether current $J^{\mu}_R$ which obeys (\ref{anomaly})
and the current $J'^{\mu}_R$ 
belonging to the supercurrent multiplet $J$ are not the same, there is a
relation \cite{LPS} between quantities associated with them
\be
r=\beta_g(1+x_g)+\beta_{ijk}x^{ijk}-\gamma_Ar^A
\label{rbeta}
\ee
where $r$ was given in Eq.~(\ref{r}).  The $r^A$ are the
non-renormalized coefficients of 
the anomalies of the Noether currents associated to the chiral
invariances of the superpotential, and --like $r$-- are strictly
one-loop quantities. The $\gamma_A$'s are linear
combinations of the anomalous dimensions of the matter fields, and
$x_g$, and $x^{ijk}$ are radiative correction quantities.
The structure of equality (\ref{rbeta}) is independent of the
renormalization scheme.

One-loop finiteness, i.e. vanishing of the $\beta$-functions at one-loop,
implies that the Yukawa couplings $\lambda_{ijk}$ must be functions of
the gauge coupling $g$. To find a similar condition to all orders it
is necessary and sufficient for the Yukawa couplings to be a formal
power series in $g$, which is solution of the REs (\ref{redeq2}).  

We can now state the theorem for all-order vanishing
$\beta$-functions.
\bigskip

\noindent{\bf Theorem:}

Consider an $N=1$ supersymmetric Yang-Mills theory, with simple gauge
group. If the following conditions are satisfied
\begin{enumerate}
\item There is no gauge anomaly.
\item The gauge $\beta$-function vanishes at one-loop
  \be
  \beta^{(1)}_g = 0 =\sum_i l(R_{i})-3\,C_{2}(G).
  \ee
\item There exist solutions of the form
  \be
  C_{ijk}=\rho_{ijk}g,~\qquad \rho_{ijk}\in\complex
  \label{soltheo}
  \ee
to the  conditions of vanishing one-loop matter fields anomalous dimensions
  \bea
  &&\gamma^{i~(1)}_j~=~0\\
  &&=\frac{1}{32\pi^2}~[ ~
  C^{ikl}\,C_{jkl}-2~g^2~C_{2}(R_{i})\delta_{ij} ].\nn
  \eea
\item these solutions are isolated and non-degenerate when considered
  as solutions of vanishing one-loop Yukawa $\beta$-functions: 
   \be
   \beta_{ijk}=0.
   \ee
\end{enumerate}
Then, each of the solutions (\ref{soltheo}) can be uniquely extended
to a formal power series in $g$, and the associated super Yang-Mills
models depend on the single coupling constant $g$ with a $\beta$
function which vanishes at all-orders.
\bigskip

It is important to note a few things:
The requirement of isolated and non-degenerate
solutions guarantees the 
existence of a unique formal power series solution to the reduction
equations.  
The vanishing of the gauge $\beta$-function at one-loop,
$\beta_g^{(1)}$, is equivalent to the 
vanishing of the R current anomaly (\ref{anomaly}).  The vanishing of
the anomalous 
dimensions at one-loop implies the vanishing of the Yukawa couplings
$\beta$-functions at that order.  It also implies the vanishing of the
chiral anomaly coefficients $r^A$.  This last property is a necessary
condition for having $\beta$ functions vanishing at all orders
\footnote{There is an alternative way to find finite theories
  \cite{strassler}.}.  

\bigskip

\noindent{\bf Proof:}

Insert $\beta_{ijk}$ as given by the REs into the
relationship (\ref{rbeta}) between the axial anomalies coefficients and
the $\beta$-functions.  Since these chiral anomalies vanish, we get
for $\beta_g$ an homogeneous equation of the form
\be
0=\beta_g(1+O(\hbar)).
\label{prooftheo}
\ee
The solution of this equation in the sense of a formal power series in
$\hbar$ is $\beta_g=0$, order by order.  Therefore, due to the
REs (\ref{redeq2}), $\beta_{ijk}=0$ too.

Thus we see that finiteness and reduction of couplings are intimately
related. 

\begin{table*}
\begin{center}\caption{The predictions of the FUT
 $SU(5)$}\label{table-fut}
\vspace{0.4cm}
\begin{tabular}{|c|c|c|c|c|c|}
\hline
$m_{\rm SUSY}$ [GeV]   &$\alpha_{3}(M_Z)$ &
$\tan \beta$  &  $M_{\rm GUT}$ [GeV]
 & $m_{b} $ [GeV]& $m_{t}$ [GeV]
\\ \hline
$200$ & $0.123$  & $53.7$ & $2.25 \times 10^{16}$
 & $5.2$ & $184.0$ \\ \hline
$500$ & $0.118$  & $54.2$ & $1.45 \times 10^{16}$
 & $5.1$ & $184.4$ \\ \hline
\end{tabular}
\end{center}
\end{table*}

\section{Finite SU(5) from reduction of couplings}

Let us next consider a realistic Finite Unified Model based on $SU(5)$. 
{}From the classification of
theories with vanishing one-loop 
$\beta$ function for the gauge coupling
\cite{HPS}, one can see that
using $SU(5)$ as gauge group there
exist only two candidate models which can 
accommodate three fermion
generations. These models contain the chiral supermutiplets
${\bf 5}~,~\overline{\bf 5}~,~{\bf 10}~,
~\overline{\bf 5}~,~{\bf 24}$
with the multiplicities $(6,9,4,1,0)$ and
 $(4,7,3,0,1)$, respectively.
Only the second one contains a ${\bf 24}$-plet which can be used
for spontaneous symmetry breaking (SSB) of $SU(5)$ down
to $SU(3)\times SU(2) \times U(1)$. (For the first model
one has to incorporate another way, such as the Wilson flux
breaking to achieve the desired SSB of $SU(5)$ \cite{kmz}).
Therefore,  we would like to concentrate only on the second model.

To simplify the situation, we neglect the intergenerational
mixing among the lepton and quark supermultiplets and consider
the following $SU(5)$ invariant cubic
superpotential for the (second)
model:
\bea
W &=& \sum_{i=1}^{3}\sum_{\alpha=1}^{4}\,[~\frac{1}{2}g_{i\alpha}^{u}
\,{\bf 10}_i
{\bf 10}_i H_{\alpha}\nn\\
& &+g_{i\alpha}^{d}\,{\bf 10}_i \overline{\bf 5}_{i}\,
\overline{H}_{\alpha}~] \nn\\
 & & +\sum_{\alpha=1}^{4}g_{\alpha}^{f}\,H_{\alpha}\, 
{\bf 24}\,\overline{H}_{\alpha}+
\frac{g^{\lambda}}{3}\,({\bf 24})^3~,~\\
&&~~~~~~~~~~~\mbox{with}~~g_{i \alpha}^{u,d}=0~\mbox{for}~i\neq \alpha~,\nn
\eea
where the ${\bf 10}_{i}$'s
and $\overline{\bf 5}_{i}$'s are the usual
three generations, and the four
$({\bf 5}+ \overline{\bf 5})$ Higgses are denoted by
 $H_{\alpha}~,~\overline{H}_{\alpha} $.
The superpotential is not the most general one, but
by virtue of the non-renormalization theorem,
this does not contradict the philosophy of 
the coupling unification by the reduction 
method (a RG invariant fine tuning is a solution
of the reduction equation). In the case at hand,
however, one can
find a discrete symmetry $Z_3\times Z_7$, together with a
multiplicative $Q$-parity, that can be imposed
on the most general cubic superpotential to arrive at the
non-intergenerational mixing \cite{kmz}.  
 
It is very interesting that demanding reduction of couplings we find a
 unique power series solution (\ref{powerser}) 
 of the dimensionless parameters of the theory in favour
 of the gauge
 coupling $g$. The unique power series solution \cite{kmz}
 corresponds to 
the Yukawa matrices
without intergenerational mixing, and yields
in the one-loop approximation
\bea
g_{t}^{2} =g_{c}^{2}~=~g_{u}^{2}&=&\frac{8}{5} g^2~,~\\
g_{b}^{2} ~=~g_{s}^{2} ~=~g_{d}^{2} &=& \frac{6}{5} g^2~,\\
g_{\tau}^{2} ~=~g_{\mu}^{2} ~=~g_{e}^{2}
{}&=&\frac{6}{5} g^2~,
\eea
where $g_i$'s stand  for the Yukawa couplings.

Moreover, the above unique solution gives in lowest order vanishing
matter fields anomalous dimensions $\gamma^{i(1)}_j$, and Yukawa
$\beta$-functions $\beta_{ijk}$.  Therefore the conditions of the
$N=1$ finiteness theorem of Chapter 5 are satisfied which in turns
guarantees the finiteness of the theory to all orders in perturbation
theory.  The point that we would like to stress here is that we have
chosen a theory with $\beta_g^{(1)}=0$ with a superpotential
restricted by some symmetries, and we have demanded reduction of
couplings.  We did not {\em impose} $\gamma^{i(1)}_j=0$.  This came
out as a bonus of the requirement of reduction of couplings
\footnote{Further details will be given in a future publication
  \cite{future}.}  and shows an intimate connection among the
reduction of couplings and finiteness in a class of supersymmetric
models with $\beta_g^{(1)}=0$.

At first sight, this GYU seems to lead
to unacceptable predictions of the fermion masses.
But this is not the case, because each generation has
an own pair of ($\overline{{\bf 5}}+{\bf 5}$)-Higgses
so that one may assume \cite{model1,kmz} that
after the diagonalization
of the Higgs fields  the effective theory is
exactly MSSM, where the pair of
its Higgs supermultiplets mainly stems from the
(${\bf 5}+\overline{{\bf 5}} $) which
couples to the third fermion generation.
(The Yukawa couplings of the first two generations
can be regarded as free parameters.)
The predictions of $M_t$ and $M_b$ for various $M_{\rm SUSY}$
are given in Table 2, where we have suppressed the threshold effects
of the superheavy as well as of the MSSM superparticles.

Adding soft
breaking terms (which are supposed not to influence the
$\beta$ functions beyond $M_{\rm GUT}$),
we can obtain supersymmetry breaking.
The conditions on
the soft breaking terms to preserve one and two-loop finiteness have been
given in refs.\cite{soft,jj}, as we have already mentioned.
It is an open problem whether there exists a suitable set of conditions
on the soft terms for all-loop finiteness.

\section{Conclusions}

As a natural extension of the unification of gauge couplings provided by 
all GUTs and the unification of Yukawa couplings, we
have introduced the idea of Gauge-Yukawa
Unification. GYU is a functional relationship among the gauge and
Yukawa couplings provided by some principle.  In our studies GYU has
been achieved by applying the principles of reduction of couplings
and finiteness. 
The consequence of GYU is that 
in the lowest order in perturbation theory
 the gauge and Yukawa couplings above  $M_{\rm GUT}$
are related  in the form
\be
g_i  = \kappa_i \,g_{\rm GUT}~,~i=1,2,3,e,\cdots,\tau,b,t~,
\label{bdry}
\ee 
where $g_i~(i=1,\cdots,t)$ stand for the gauge 
and Yukawa couplings, $g_{\rm GUT}$ is the unified coupling,
and
we have neglected  the Cabibbo-Kobayashi-Maskawa mixing 
of the quarks.
 So, Eq.~(\ref{bdry}) exhibits a set of boundary conditions on the 
the renormalization group evolution for the effective theory
below $M_{\rm GUT}$, which we have assumed to 
be the MSSM. 
We have shown \cite{kmoz2} that it is
possible to construct some 
supersymmetric GUTs with GYU in the 
third generation that can
 predict the bottom and top
quark masses in accordance with the recent experimental data. 
This means that the top-bottom hierarchy 
could be
explained in these models,
 in a similar way as 
the hierarchy of the gauge couplings of the SM
can be explained if one assumes  the existence of a unifying
gauge symmetry at $M_{\rm GUT}$. 

 \begin{figure}[t]
  \begin{center}
    \leavevmode
    \rotate[r]{
      \mbox{
        \epsfysize=7cm
        \epsffile{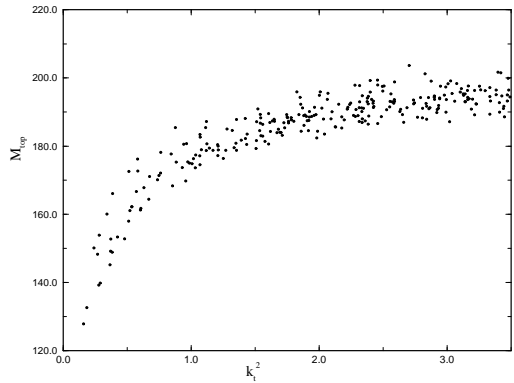}
        }
      }
  \end{center}
  \caption[\ ]{The dependence of the top mass $M_t$ with $k^2_t$, at
    fixed $M_{SUSY}=500$ GeV.
    As we can see, after $k^2_t\sim 2.0$ the top mass goes to its infrared
    fixed point value.}
  \label{fig:3}
\end{figure}

It is clear that the GYU scenario  is the most predictive scheme as far as
the mass of the top quark is concerned.
It may be worth recalling the predictions for $M_t$
of ordinary GUTs, in particular of supersymmetric $SU(5)$ and
$SO(10)$.  The MSSM with $SU(5)$ Yukawa boundary unification allows
$M_t$ to be anywhere in the interval between 100-200 GeV 
for varying $\tan \beta$, which is now a free parameter.  Similarly,
the MSSM with $SO(10)$ Yukawa 
boundary conditions, {\em i.e.} $t-b-\tau$ Yukawa Unification, gives
$M_t$ in the interval 160-200 GeV. 
We have analyzed \cite{kmoz2} the infrared quasi-fixed-point behaviour of
the $M_t$ prediction in some detail. In particular we have seen that
the {\em infrared value} for large $\tan \beta$ depends on  $\tan
\beta$  and its lowest value is $\sim 188$ GeV. 
 Comparing this with
the experimental value $m_t = (176.8 \pm 6.5)$ GeV we may conclude 
that the present data on $M_t$ cannot be explained from the infrared
quasi-fixed-point behaviour alone (see Figure 1).

Clearly, to exclude or verify different GYU models,
 the experimental as well as theoretical uncertainties
have to be further reduced.
One of the largest theoretical uncertainties 
 in FUT  results
from the not-yet-calculated threshold effects 
of the superheavy particles.
Since the structure of  the superheavy 
particles is basically fixed,
 it will be possible to
bring these threshold effects under control,
which will  reduce the uncertainty of 
the $M_t$ prediction \cite{kmoz2}.
There we have been regarding the MSSM threshold correction
 $\delta^{\rm MSSM} M_t$  
as unknown because we do not have 
sufficient information on the superpartner spectra.
Recently, however, we have demonstrated \cite{kmz-pert} how to extend
 the principle of reduction of couplings in a way as to include the
 dimensionfull parameters.  As a result, it is in principle possible
 to predict the superpartner spectra as well as the rest of the
 massive parameters of a theory. 

\section*{Acknowledgements}

It is a pleasure for one of us (G.Z) to thank the
Organizing Committee for the warm hospitality.

\end{document}